%% file: 0-paper.tex
\documentclass[sigconf]{acmart}
\usepackage{fancyhdr}
\usepackage{graphicx}
\usepackage{tabularx}
\usepackage{subcaption}
\usepackage{textcomp}
\usepackage{siunitx}
\usepackage{amsmath, amsthm}
\usepackage{graphbox}
\usepackage{multirow}
\usepackage{makecell}
\usepackage{fixltx2e}
\usepackage{setspace}
\usepackage{booktabs}
\usepackage{array}
\usepackage{tabularx}
\usepackage{calc}
\usepackage{makecell}
\usepackage[T1]{fontenc}
\usepackage[utf8]{inputenc}
\usepackage[font=small,labelfont=bf]{caption}
\usepackage{booktabs}
\usepackage{subcaption}
\usepackage{enumitem}
\usepackage{multirow}
\usepackage{threeparttable}
\usepackage{makecell}
\usepackage{hhline}

\AtBeginDocument{%
	\providecommand\BibTeX{{%
			\normalfont B\kern-0.5em{\scshape i\kern-0.25em b}\kern-0.8em\TeX}}}

\setcopyright{acmcopyright}
\copyrightyear{2018}
\acmYear{2018}
\acmDOI{XXXXXXX.XXXXXXX}

\acmConference[Conference acronym 'XX]{Make sure to enter the correct
	conference title from your rights confirmation emai}{June 03--05,
	2018}{Woodstock, NY}
\acmPrice{15.00}
\acmISBN{978-1-4503-XXXX-X/18/06}

\begin{document}
	\begin{sloppypar}
		
	\title{Hybrid Audio Detection Using Fine-Tuned Audio Spectrogram Transformers: A Dataset-Driven Evaluation of Mixed AI-Human Speech}

		\author{Kunyang Huang$^\ast$, Bin Hu$^\ast$}
	\affiliation{
		\institution{$^\ast$Kean University}
		\city{}
		\state{}
		\country{}}
		\renewcommand{\shortauthors}{Kunyang et al.}
	\input{abstract}


\maketitle
    \input{01-introduction}
    \input{02-Relatedwork}
	\input{03-Datasets}
    \input{04-Methodology}
	\input{05-Experiment}

    \input{06-Conclusion}
	%



	\appendix
	
\end{sloppypar}

\end{document}

%% file: abstract.tex
\begin{abstract}

The rapid advancement of artificial intelligence (AI) has enabled sophisticated audio generation and voice cloning technologies, posing significant security risks for applications reliant on voice authentication. While existing datasets and models primarily focus on distinguishing between human and fully synthetic speech, real-world attacks often involve audio that combines both genuine and cloned segments. To address this gap, we construct a novel hybrid audio dataset incorporating human, AI-generated, cloned, and mixed audio samples. We further propose fine-tuned Audio Spectrogram Transformer (AST)-based models tailored for detecting these complex acoustic patterns. Extensive experiments demonstrate that our approach significantly outperforms existing baselines in mixed-audio detection, achieving 97\% classification accuracy. Our findings highlight the importance of hybrid datasets and tailored models in advancing the robustness of speech-based authentication systems.

\end{abstract}

%% file: 01-introduction.tex
\begin{figure}
    \centering
    \includegraphics[width=1\linewidth]{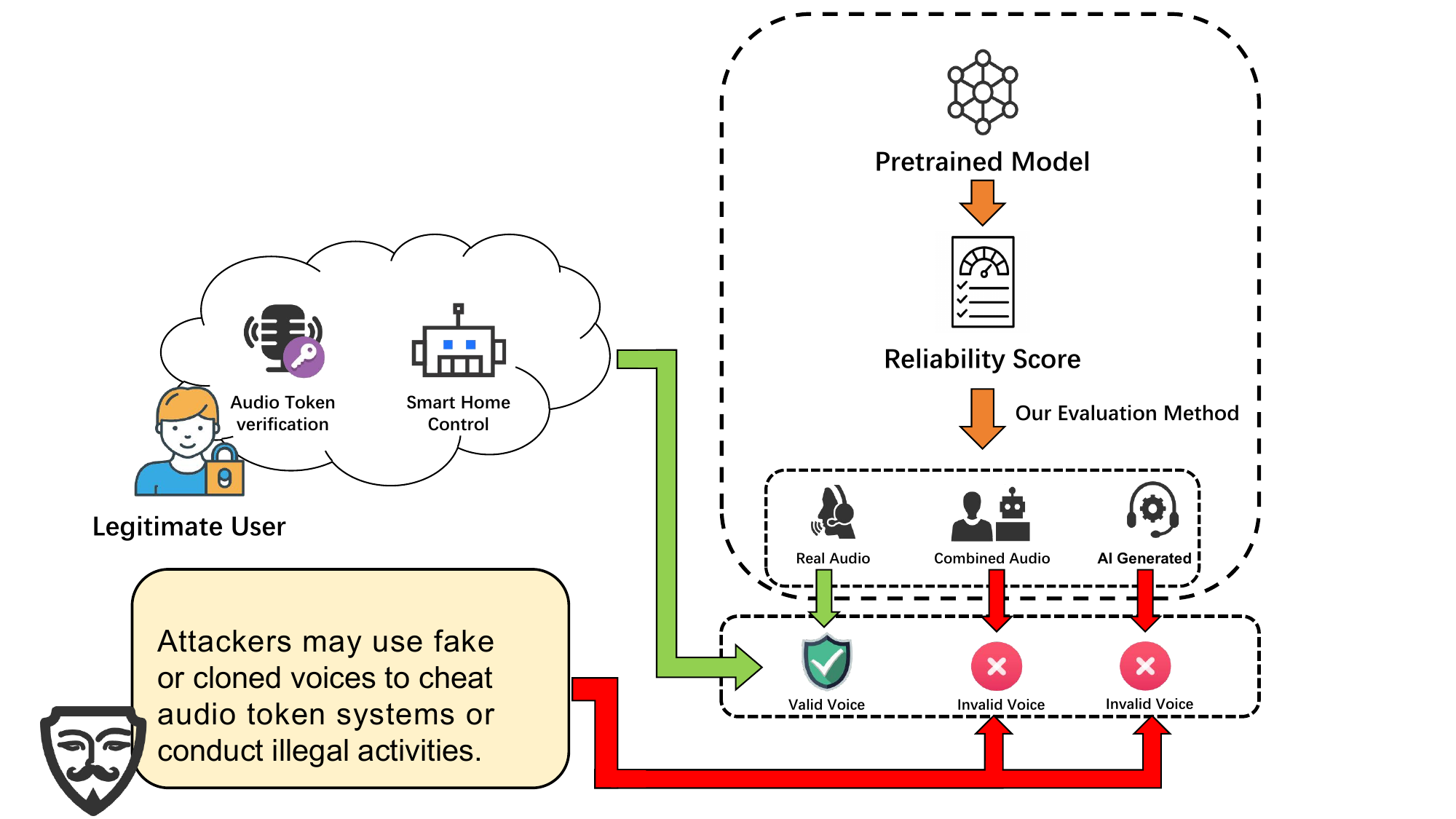}
    \caption{Our evaluation framework in a smart home audio token system. It effectively prevents unauthorized access by detecting non-authentic voices, including AI-synthesized, AI-cloned, and mixed human-AI speech, thereby ensuring secure control over smart home environments.}

    \label{fig:audio-token-system}
\end{figure}
\section{Introduction}

Voice biometrics have witnessed rapid adoption for user authentication, particularly during the COVID-19 pandemic, as social distancing protocols accelerated the shift toward contactless security systems. Compared to password-based and fingerprint-based authentication, voice-based authentication offers greater efficiency, user convenience, and cost-effectiveness~\cite{Ibrar2023}. Furthermore, speech interaction technologies have revolutionized human-computer interaction paradigms, enabling transformative applications in smart homes, financial services, and medical education via intelligent voice assistants and authentication systems~\cite{Kim2016}.

However, while enhancing accessibility and operational efficiency, these advancements have introduced critical cybersecurity vulnerabilities. The proliferation of artificial intelligence (AI) technologies has enabled the creation of highly realistic synthetic audio and cloned voices~\cite{Liu2023}, which malicious actors increasingly exploit to impersonate trusted individuals. For example, in a widely publicized incident in early 2020, a Hong Kong bank manager authorized a \$35 million transfer based on an AI-cloned voice~\cite{Brewster2021}, illustrating the devastating potential of such attacks. This emerging threat landscape necessitates the development of robust anti-spoofing mechanisms capable of detecting sophisticated synthetic media.

To counter these risks, significant research efforts have been dedicated to detecting AI-generated speech. Automatic speaker verification (ASV) systems~\cite{Lei2022, Chen2024, Liu2023} have become the predominant technology for voice authentication in mobile phones, smart speakers, and call centers~\cite{Ibrar2023}. ASV systems typically rely on the extraction of distinctive vocal features—such as pitch, formants, spectral characteristics, and phase information—which are compared against stored biometric voiceprints to verify speaker identity~\cite{Chouchane2024}. A close feature match grants access, while significant discrepancies trigger security responses or secondary authentication measures~\cite{VS2024}.

Early countermeasure systems primarily utilized hand-crafted acoustic features, such as Constant-Q Cepstral Coefficients (CQCC)~\cite{Todisco2017cqcc} combined with Gaussian Mixture Models (GMMs), for spoof detection. Although effective against specific attack types, these systems often lacked generalization when faced with previously unseen spoofing methods or recording conditions.

With the rise of deep learning, more robust methods were introduced. Convolutional Neural Networks (CNNs)~\cite{Zhang2021cnnspoof} and Deep Neural Networks (DNNs)~\cite{Alam2021asvspoof} improved detection accuracy by automatically learning discriminative features from spectrogram representations. Attention-based architectures, such as the Audio Spectrogram Transformer (AST)~\cite{gong2021astaudiospectrogramtransformer}, further enhanced performance by modeling global time-frequency dependencies. Self-supervised learning paradigms, including SSAST~\cite{goel2024towards}, also demonstrated notable improvements in spoof detection under limited labeled data conditions.

Despite these advancements, current detection systems still suffer from several limitations rooted in their training data. Existing public datasets such as FoR-Original~\cite{reimao2019for}, ASVspoof 2015/2019/2021~\cite{nautsch2021asvspoof}, and others focus predominantly on binary classification—distinguishing between genuine and fully spoofed utterances. While ASVspoof datasets comprehensively cover speech synthesis, voice conversion, and replay attacks, they do not model hybrid attacks where genuine and synthetic speech are interleaved within a single utterance. Similarly, datasets like HAD~\cite{yi2021half} and ADD2023-PF~\cite{yi2023add} introduced tampered audios with localized modifications but still emphasized isolated segment replacements rather than complex multi-source compositions. Moreover, many of these datasets lack diversity in speaker demographics, consistent utterance lengths, and precise annotations of tampering types and positions.

Consequently, models trained on these datasets tend to overestimate their real-world performance and are often ill-equipped to handle hybrid and partial spoofing attacks~\cite{Evans2015, Vestman2019}. Furthermore, assumptions of fixed-length speech, minimal background noise, and perfect recording conditions further limit their applicability to noisy, unconstrained real-world environments.

To address these critical gaps, this research introduces a novel hybrid speech dataset and evaluation framework. Specifically, twelve participants aged 19--36 years recorded 104 linguistically diverse sentences, spanning structured content, conversational speech, and grammatically anomalous constructions. Using these authentic recordings, we generated cloned and AI-synthesized variants, subsequently constructing seven distinct hybrid audio groups representing various combinations of human, cloned, and synthetic speech. Each sample was meticulously annotated with speaker metadata, tampering patterns, and authenticity labels.

Unlike previous datasets, our corpus emphasizes:
\begin{itemize}
    \item The creation of composite audios combining multiple spoofing types within a single utterance.
    \item Fine-grained annotations capturing tampering regions, source modalities, and demographic variations.
    \item Consistent utterance lengths to facilitate hybrid segment detection and long-form speech analysis.
\end{itemize}

To evaluate the discriminative capabilities of detection models on this hybrid dataset, we fine-tuned two variants of the Audio Spectrogram Transformer (AST) architecture~\cite{mit_classification, gong2021astaudiospectrogramtransformer}. By systematically categorizing and classifying the audio samples, we assess model performance under complex and realistic conditions, elucidating the challenges and potential of hybrid audio detection.

The main contributions of this study are summarized as follows:
\begin{itemize}
    \item We construct a comprehensive dataset encompassing human, AI-generated, cloned, and mixed-source audio samples, significantly enriching the resource pool for research on voice anti-spoofing.
    \item We systematically vary the degree of audio cloning and hybridization to evaluate model performance across different complexity levels, enabling a deeper understanding of detection capabilities and limitations.
    \item We propose two fine-tuned AST-based models tailored for hybrid speech detection, demonstrating enhanced accuracy and robustness under realistic adversarial conditions.
\end{itemize}

Collectively, these contributions advance the development of reliable and robust speech-recognition systems capable of countering emerging threats in cybersecurity-critical applications.

%% file: 02-Relatedwork.tex
\vspace{-0.3cm}
\section{Related Work}
\subsection{Deepfake Audio Detection Models}
Early efforts in deepfake speech detection primarily relied on traditional machine learning models, such as Gaussian Mixture Models (GMM)~\cite{wen2022multi}, which achieved competitive performance on benchmarks like ASVspoof 2021 Logical Access (LA). However, their limited capacity to model global temporal patterns rendered them ineffective against sophisticated attacks such as replay-based spoofing.

Subsequent approaches leveraged deep learning architectures, notably Light Convolutional Neural Networks (LFCNN)~\cite{nautsch2021asvspoof} and Long Short-Term Memory (LSTM) networks~\cite{cheng2023analysis}, to improve feature extraction via end-to-end learning. Despite achieving enhanced detection accuracy compared to traditional methods, these models often relied on local or sequential operations, hindering their ability to generalize across diverse spoofing types and unconstrained recording conditions.

A significant breakthrough came with the introduction of the Audio Spectrogram Transformer (AST)~\cite{gong2021astaudiospectrogramtransformer}, which treats audio spectrograms as two-dimensional images and applies self-attention to model global time-frequency dependencies. AST achieved 95.6\% accuracy on the ESC-50 environmental sound classification task and 97.4\% accuracy on Speech Commands V2, demonstrating strong generalization and feature modeling capabilities.

Building upon AST, self-supervised learning frameworks like SSAST-CL~\cite{goel2024towards} achieved further improvements. By leveraging contrastive pretraining, SSAST-CL achieved an Equal Error Rate (EER) of 4.74\% on ASVspoof 2021 LA, surpassing many supervised baselines and highlighting the potential of unlabeled data for spoofing detection.

Parallel research explored joint optimization of Automatic Speaker Verification (ASV) and Countermeasure (CM) systems. Kanervisto et al.~\cite{kanervisto2022optimizing} proposed reinforcement learning strategies to directly minimize tandem Detection Cost Function (t-DCF) metrics, achieving a 20\% relative reduction in t-DCF compared to independently fine-tuned baselines. This joint optimization approach improved robustness against hybrid and hard-to-detect attack classes (e.g., A17--A19).

Recent advancements also incorporated physical acoustic cues to enhance detection reliability. For instance, micro-signature modeling of microphone imperfection patterns~\cite{patil2022microsignatures} and spectral-temporal modulation analysis~\cite{cheng2023analysis} demonstrated improved spoofing detection, particularly under replay and replay-enhanced attacks.

While these innovations significantly advanced the field, challenges remain. Most existing models assume uniform speech types (genuine or spoofed) within a single utterance and are rarely evaluated under realistic conditions involving hybrid or mixed-source audios. Improving detection systems' ability to handle hybrid attacks, multi-stage forgeries, and noisy environmental factors remains a critical direction for future work.

\subsection{Spoofed Audio Datasets}
Progress in deepfake speech detection has been tightly coupled with the availability of specialized datasets. The FoR-Original dataset~\cite{reimao2019for} provided early resources for synthetic speech detection by offering bona fide and TTS-generated utterances. However, its binary nature (genuine vs. spoofed) limited its ability to model complex attack strategies encountered in practical scenarios~\cite{li2024audio}.

The ASVspoof corpora (2015, 2019, 2021)~\cite{nautsch2021asvspoof} represented substantial advancements, introducing diverse spoofing modalities such as speech synthesis, voice conversion, and replay attacks across logical and physical access scenarios. Nevertheless, these datasets primarily focus on full-utterance spoofing, lacking examples where genuine and synthetic content are intertwined within a single audio file—a growing concern for adversarial attack realism.

Efforts to model partial forgeries have emerged with the HAD dataset~\cite{yi2021half}, which introduced utterances with isolated word-level replacements using TTS systems. While beneficial for fine-grained tampering detection, HAD's scope is constrained to word substitution and short utterances, limiting its suitability for evaluating long-form or conversation-level spoofing attacks. The ADD2023-PF dataset~\cite{yi2023add} further addressed partial tampering by providing segment-level annotations of manipulated regions within composite audios. However, the lack of detailed documentation regarding spoofing methods and the inconsistency of attack types across samples pose challenges for building robust, generalizable models. Additional efforts, such as partially fake audio datasets~\cite{alam2022partial} and adversarially perturbed speech corpora~\cite{wu2020adversarial}, have introduced valuable resources for evaluating resilience to adversarial attacks. Nevertheless, many existing datasets still suffer from major limitations: they lack hybrid samples combining multiple spoofing techniques within the same utterance, they offer insufficient fine-grained annotations describing tampering types and boundaries, and they fail to maintain consistent utterance lengths necessary for real-world deployment.

To address these deficiencies, our proposed dataset introduces composite audios combining human, cloned, and AI-generated speech in varied proportions. It incorporates detailed metadata including speaker demographics, spoofing methods, and segment-level boundaries, while ensuring standardized utterance durations to facilitate both classification and localization tasks. This design simulates complex real-world adversarial conditions more accurately and supports the development of next-generation hybrid speech detection systems capable of operating under noisy, unconstrained environments.

%% file: 03-Datasets.tex
\vspace{-0.2cm}
\section{Dataset Construction}
\label{sec:dataset-construction}

\subsection{Design Policy}

The proposed hybrid anti-spoofing audio dataset is designed to evaluate detection systems under both controlled and adversarial conditions. Motivated by the increasing complexity of audio forgery scenarios in real-world applications, this dataset incorporates clean genuine audio, cloned and AI-generated speech, hybrid utterances, and adversarial perturbations including noise, codec compression, and channel degradation. Our goal is to assess the generalization and robustness of spoof detection models under challenging, unseen, and mixed-modality scenarios.

The dataset consists of two major versions:
\begin{itemize}
\item \textbf{Clean Hybrid Version:} Includes genuine, cloned, synthetic, and hybrid (human-synthetic mix) speech in controlled recording conditions.
\item \textbf{Noisy/Degraded Version:} Applies various distortions (noise, compression, and filtering) to clean audio to simulate real-world audio environments.
\end{itemize}

Each version is split into training, development, and test subsets, with the test subset further divided into seen and unseen attack types for robust generalization evaluation.

\subsection{Clean Real Audio Collection}

Real human speech was collected from twelve participants (aged 19–38, balanced across gender) in an acoustically controlled indoor setting using studio-grade USB microphones. Audio was recorded at 44.1 kHz and downsampled to 16 kHz for consistency with existing spoofing corpora.

Each speaker recorded 104 scripted sentences across 8 linguistically diverse categories to ensure phonetic, syntactic, and semantic variation:
\begin{itemize}
\item Alphanumeric combinations
\item Pure alphabetic strings
\item Numerical sequences
\item Natural English phrases (New Concept English)
\item Semantically coherent sentence pairs
\item Semantically unrelated sentence pairs
\item Grammatically incorrect sentences
\item Complex anomaly-infused constructions
\end{itemize}

These categories were chosen to emulate both structured and spontaneous speech patterns, supporting tasks such as forgery detection, localization, and contextual spoof analysis.
\begin{itemize}
    \item Structured content (e.g., alphanumeric sequences, pure alphabetic sequences)
    \item Natural language sentences (e.g., coherent vs. unrelated sentences)
    \item Complex grammatical constructions (e.g., sentences with embedded grammatical errors or semantic anomalies)
\end{itemize}

This design ensures the dataset captures phonetic variability, syntactic complexity, and spontaneous speech phenomena crucial for training generalizable models. Table~\ref{tab:model_performance} summarizes the sentence composition.

\begin{table}[ht]
\centering
\resizebox{\linewidth}{!}{%
\begin{tabular}{@{}lr@{}}
\toprule
\textbf{Type Name} & \textbf{Number of Sentences} \\ 
\midrule
Alphanumeric Combination Sentences & 8 \\
Pure Alphabetic Sentences & 8 \\
Numeric Sequence Sentences & 8 \\
New Concept English 2 Lesson 1 Sentences & 16 \\
Semantically Coherent Sentence Pairs & 16 \\
Semantically Unrelated Sentences & 16 \\
Grammatical Error-Embedded Sentences & 16 \\
Complex Sentences with Semantic-Grammatical Anomalies & 16 \\
\bottomrule
\end{tabular}
}
\caption{Composition of sentence types in the human speech dataset.}
\label{tab:model_performance}
\end{table}

\subsection{Cloned and Synthetic Audio Generation}

We used a Tacotron 2-based TTS system equipped with speaker embeddings to synthesize cloned speech. Four levels of reference embedding conditions were implemented to systematically vary the fidelity of voice cloning:
\begin{enumerate}[label=C$_{\arabic*}$:]
\item Single sentence reference (minimal context)
\item Subset corpus reference (16-sentence context)
\item Full speaker corpus (104 sentences)
\item Target sentence embedding (maximum context)
\end{enumerate}

Each cloned sample was evaluated using the ERes2NetV2 model~\cite{chen2024eres2netv2} to measure speaker similarity. All cloning conditions exceeded a 0.70 reliability score, validating their effectiveness in simulating believable spoof attacks.

In addition to cloned samples, fully synthetic speech was generated using zero-shot TTS systems, further enriching the diversity of spoofing artifacts. Each synthesized audio was paired with a corresponding transcript to ensure semantic validity.

\subsection{Hybrid Audio Construction}

Hybrid utterances were crafted by concatenating genuine and synthetic segments within a single audio stream. This mimics practical spoofing cases where attackers inject synthesized phrases into authentic recordings.

Three composition patterns were utilized:
\begin{itemize}
\item Human speech followed by AI-speech (H$\rightarrow$S)
\item AI-speech followed by human speech (S$\rightarrow$H)
\item Interleaved segments (H$\leftrightarrow$S$\leftrightarrow$H)
\end{itemize}

Transitions were smoothed using 10 ms cross-fading to eliminate audible artifacts. Each file was annotated with source types, gender, segment arrangement, and spoof type (0 = real, 1 = fake).

\subsection{Noisy and Codec-Augmented Dataset}

To simulate real-world variability, we created a noisy version of the dataset by introducing additive distortions across three dimensions:

\textbf{(a) Additive Noise:} Background noise (e.g., street sounds, café ambiance, white noise) was sourced from open noise datasets and injected at SNRs of 10, 15, 20, and 30 dB.

\textbf{(b) Channel Simulation:} Low-pass filtering (4 kHz cutoff) and spectral shaping were applied to simulate recording over mismatched devices.

\textbf{(c) Compression Artifacts:} Audio was encoded and decoded using the Opus codec at 16 and 24 kbps, introducing quantization noise and frame loss.

Each corrupted utterance retained its original label, but was further annotated with SNR levels, noise type, and degradation parameters to support domain-aware training.

\subsection{Dataset Statistics}

The final dataset consists of two major partitions:
\begin{itemize}
\item \textbf{Hybrid Clean Dataset:} 1,248 utterances (312 per class × 4 classes: human, cloned, AI-generated, hybrid).
\item \textbf{Noisy/Compressed Dataset:} 1,248 utterances derived from the clean set with added distortions.
\end{itemize}

Each speaker contributed equally across all categories, ensuring demographic and linguistic balance.

\subsection{Design Advantages}

The dataset offers several innovations over prior benchmarks such as ASVspoof 2019~\cite{nautsch2021asvspoof}, HAD~\cite{yi2021half}, and ADD2023-PF~\cite{yi2023add}:

\begin{itemize}
\item \textbf{Multi-source Hybrid Samples:} Unlike binary spoof corpora, we provide realistic adversarial combinations in a single utterance.
\item \textbf{Multi-fidelity Cloning:} Four levels of speaker context allow controlled evaluation of cloning accuracy and spoofing difficulty.
\item \textbf{Domain Transfer Evaluation:} Noisy and degraded samples simulate mobile devices, low-bandwidth transmission, and user environments.
\item \textbf{Rich Metadata:} Each sample includes spoof origin, segment order, gender, and degradation type to support interpretability and explainability research.
\end{itemize}

In conclusion, our dataset fills critical gaps in existing benchmarks by introducing hybrid compositions, variable fidelity spoofing, and practical noise artifacts—supporting future research in robust, real-world speech anti-spoofing.

\subsection{Dataset}

The dataset comprises two major partitions:

\begin{itemize}
    \item \textbf{Clean Hybrid Dataset:} 1,248 utterances across four balanced spoofing classes, constructed using speaker-controlled human recordings and Tacotron 2-based synthesis.
    \item \textbf{Noisy/Degraded Dataset:} 1,248 utterances derived from the clean version by applying real-world distortions, including additive noise (at multiple SNR levels), low-pass filtering, and Opus-based codec compression.
\end{itemize}

Each utterance is annotated with spoofing type, speaker ID, gender, and distortion parameters (if applicable). The dataset design emphasizes diversity in spoofing strategies, segment arrangements, and degradation conditions.

\begin{figure}[H]
    \centering
    \includegraphics[width=1\linewidth]{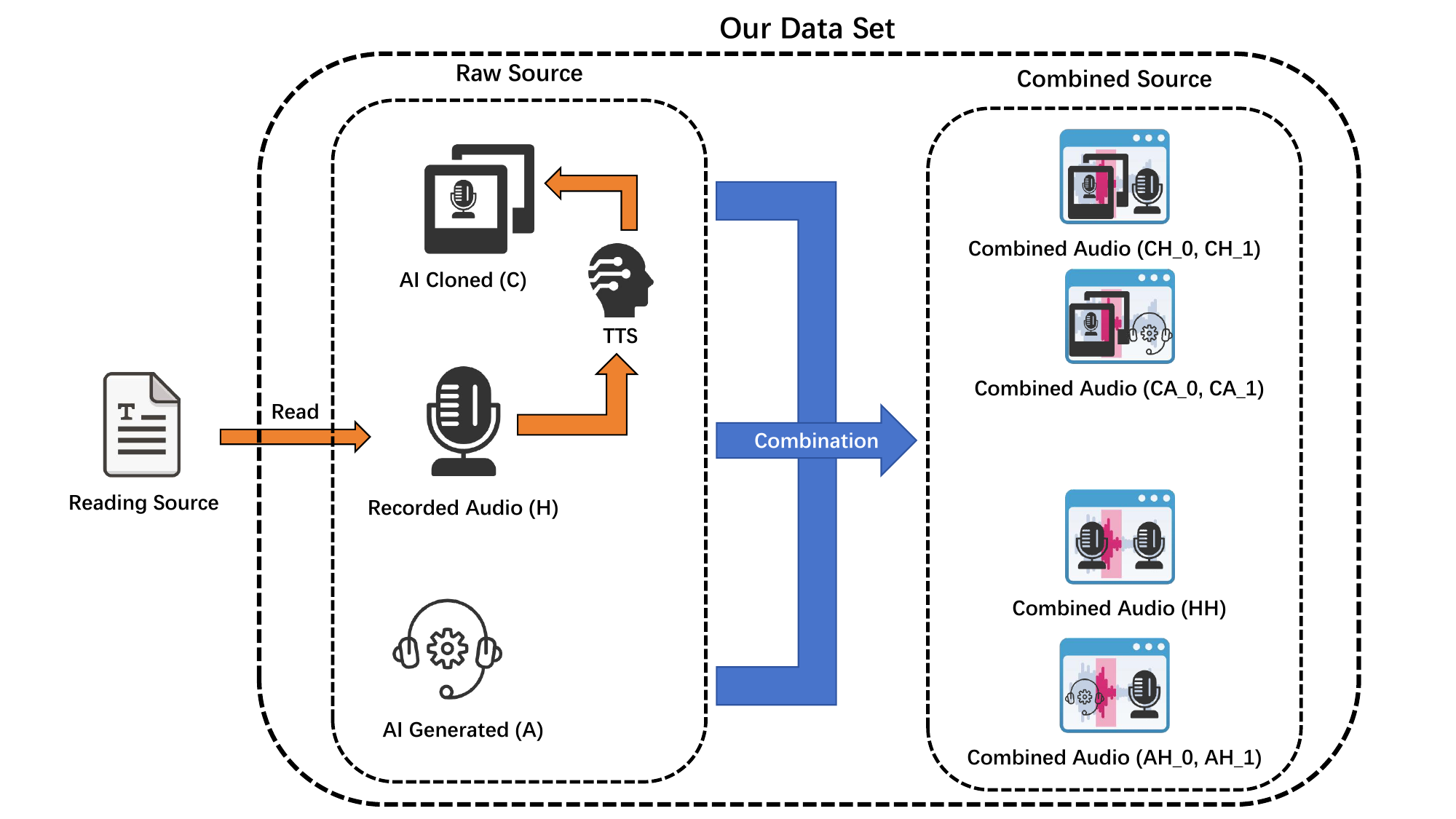}
    \caption{Construction process of the hybrid sentence dataset, encompassing recording, cloning, synthesis, and composite generation.}
    \label{fig:dataset-framework}
\end{figure}

%% file: 04-Methodology.tex
\section{Detection Methodologies}

\subsection{Overview}

This study constructs a comprehensive hybrid audio dataset to support the detection of adversarial speech manipulations, including speech synthesis, voice conversion, and replay attacks~\cite{Ibrar2023}. To replicate real-world complexity, we developed six hybrid configurations by combining these attack modalities (e.g., synthetic-replay, clone-replay). Cloned samples were generated using state-of-the-art TTS systems, and replay conditions were simulated via intra-speaker temporal concatenation.

To evaluate detection performance, we employed two high-impact variants of the Audio Spectrogram Transformer (AST)~\cite{gong2021astaudiospectrogramtransformer}, selected from the Hugging Face model hub for their transfer learning readiness and practical deployment coverage:

\begin{itemize}
\item \textbf{MIT/ast-finetuned-audioset-10-10-0.4593:} A general-purpose AST model pretrained on 10-second audio clips from AudioSet, encompassing diverse sound classes such as speech, music, and ambient noise. This model provides broad acoustic coverage and is ideal for assessing generalizability to unseen spoofing patterns.

\item \textbf{MattyB95/AST-ASVspoof2019-Synthetic-Voice-Detection:} A domain-specific AST model fine-tuned on the ASVspoof 2019 LA dataset. It focuses on synthetic voice detection and serves as a specialized baseline for binary spoof classification.
\end{itemize}

These complementary models enable an analysis of both broad-domain robustness and task-specific optimization under hybrid and adversarial spoofing conditions.

The Audio Spectrogram Transformer (AST) is a convolution-free architecture based on the Vision Transformer (ViT) paradigm, designed for end-to-end audio classification using spectrogram representations. AST directly models the long-range time-frequency dependencies in audio signals using self-attention mechanisms.

\begin{figure}[H]
    \centering
    \includegraphics[width=1\linewidth]{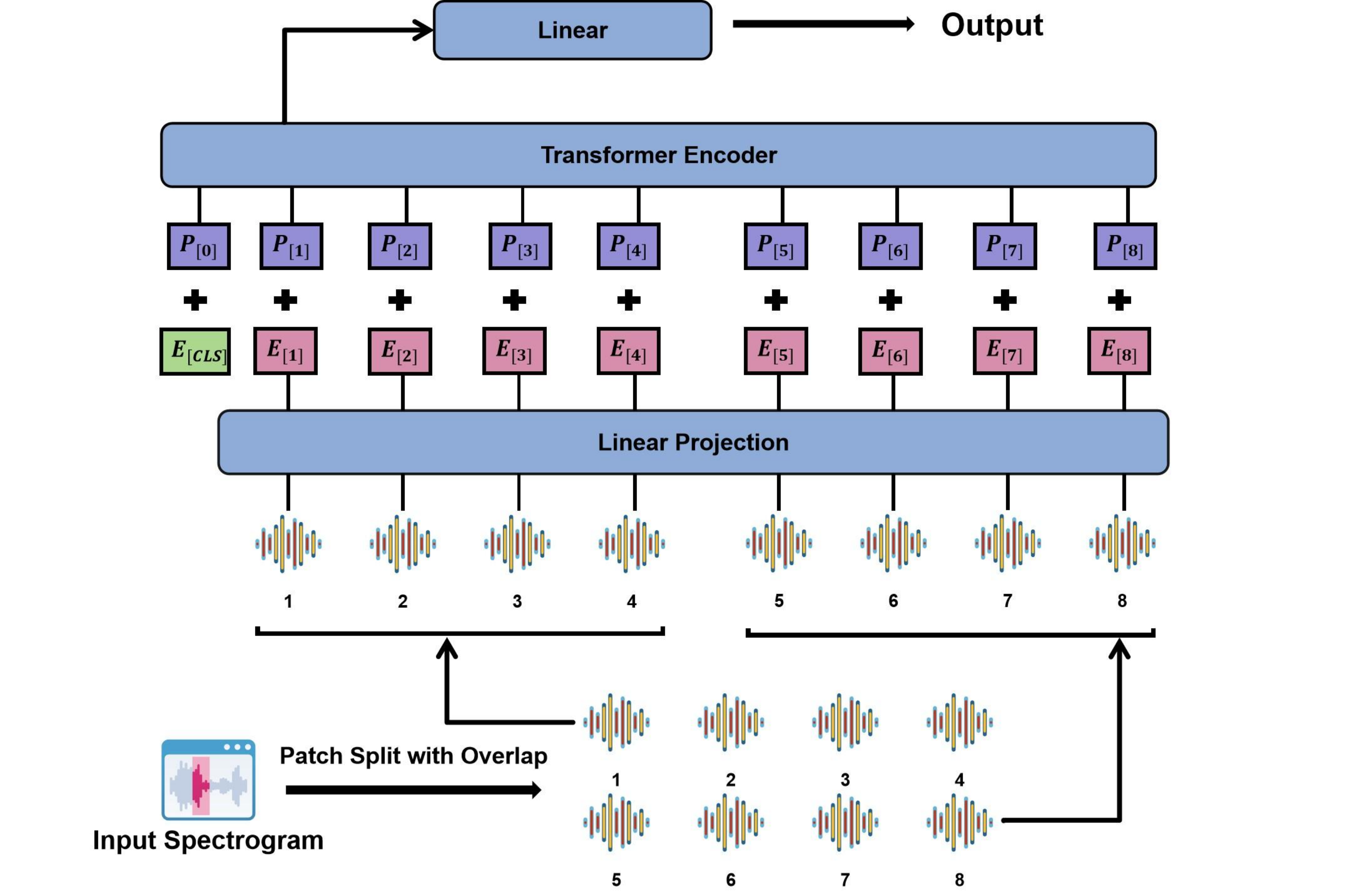}
    \caption{Architectural framework of the baseline Audio Spectrogram Transformer (AST) model.}
    \label{fig:ast-architecture}
\end{figure}

\subsection{Model Architecture: Audio Spectrogram Transformer (AST)}

\subsubsection{Design Overview}

The AST architecture adapts the Vision Transformer (ViT) paradigm for audio spectrogram classification. Rather than relying on convolutional filters, AST treats audio spectrograms as visual patches, enabling the model to learn long-range time-frequency dependencies from early layers. The design is intentionally modular, consisting of three key stages:

\begin{enumerate}
\item \textbf{Spectrogram Conversion:} Raw waveforms are first converted into 128-bin log-Mel spectrograms using a 25 ms Hamming window and 10 ms hop length.

\item \textbf{Patch Embedding:} The spectrogram is segmented into overlapping \(16 \times 16\) patches with 6-frame overlaps in both frequency and time. These patches are flattened and linearly projected to a 768-dimensional space. A trainable position embedding is added to encode spatial structure. A learnable [CLS] token is prepended to aggregate sequence-level representation.

\item \textbf{Transformer Encoding and Classification:} The patch sequence is passed through 12 Transformer encoder blocks, each containing multi-head self-attention (12 heads), layer normalization, residual connections, and GELU-activated feed-forward layers. The final [CLS] token output serves as the global representation, which is classified into 4 spoofing categories via a softmax head.
\end{enumerate}

This architecture benefits from high scalability, flexibility across input lengths, and strong inductive bias for capturing hybrid audio characteristics. Unlike CNNs, which emphasize localized feature extraction, AST excels at modeling distributed spoofing artifacts across entire utterances.

\subsubsection{Architectural Details}

The input to the AST model is a spectrogram of shape $128 \times T$, where $T$ is determined by the duration of the audio. Key architectural components include:

\begin{itemize}
    \item \textbf{Patch Size:} $16 \times 16$ with overlap
    \item \textbf{Embedding Dimension:} 768
    \item \textbf{Transformer Layers:} 12 encoder blocks
    \item \textbf{Attention Heads:} 12 per encoder
    \item \textbf{Position Encoding:} Trainable vectors for patch indexing
    \item \textbf{Classification Token:} [CLS] prepended to the sequence
\end{itemize}

The final classifier outputs one of four class labels:

\begin{itemize}
    \item Class 0 – Human (bona fide) speech
    \item Class 1 – Cloned (speaker-matched) synthetic speech
    \item Class 2 – AI-generated synthetic speech
    \item Class 3 – Hybrid audio (mixed human and synthetic segments)
\end{itemize}

\subsubsection{Transfer Learning and Optimization Strategy}

We initialized both AST variants using pretrained weights: the MIT model was trained on AudioSet (over 2M weakly labeled clips), and MattyB95 was fine-tuned on ASVspoof 2019. We then re-fine-tuned both models on our hybrid spoofing dataset to maximize performance under realistic and adversarial conditions.

Training configuration includes:

\begin{itemize}
    \item Optimizer: Adam with weight decay
    \item Learning rate: $2 \times 10^{-5}$ with cosine decay
    \item Batch size: 3
    \item Epochs: 20 with early stopping on validation loss
    \item Input: Zero-padded log-Mel spectrograms of 6 seconds
\end{itemize}

This strategy not only accelerates convergence but also enables the models to maintain discriminative power in noisy, codec-compressed, or mixed-source audio—critical for real-world ASV systems.

\subsubsection{Pretraining Strategy and ViT Adaptation}

A major advantage of AST is its ability to benefit from large-scale visual pretraining, owing to the structural similarity between audio spectrograms and images. However, due to the data-hungry nature of Transformers, training AST from scratch is infeasible for most audio tasks. We therefore employ cross-modal transfer learning from Vision Transformers (ViT), adapting pretrained weights from ImageNet classification.

To enable this transfer, several adjustments are made:

\begin{itemize}
    \item \textbf{Channel Adaptation:} ViT expects 3-channel RGB input, while AST takes single-channel spectrograms. We average the weights of ViT's input channels to initialize AST's patch embedding layer, simulating a triplicated mono-channel.
    
    \item \textbf{Input Normalization:} Spectrograms are normalized to zero mean and standard deviation 0.5 to match ViT initialization statistics.
    
    \item \textbf{Positional Embedding Resizing:} ViT uses fixed-size positional embeddings (e.g., $24 \times 24$ for 384×384 images), while audio inputs vary in length. We adopt a cut-and-bilinear interpolation strategy to reshape ViT’s 2D positional embeddings to AST’s patch grid (e.g., $12 \times 100$ for 10-second audio). The [CLS] token’s positional vector is reused without change.
    
    \item \textbf{Classification Layer Reset:} The final ViT classification layer is replaced with a new 4-class head corresponding to human, cloned, AI-generated, and hybrid audio.
\end{itemize}

In this work, we specifically adapt weights from a pretrained DeiT (Data-efficient Image Transformer) model~\cite{touvron2021training}, which was trained on ImageNet using knowledge distillation. This model contains 87M parameters and achieves 85.2\% top-1 accuracy. To align with AST, we average its dual [CLS] tokens and discard its image-specific output layer.

\subsubsection{Training Configuration and Label Design}

AST was fine-tuned using speaker-disjoint partitions from our hybrid dataset. The input was zero-padded or cropped to 6-second segments and transformed into log-Mel spectrograms (128 bins, 25ms frame, 10ms hop). Training used the Adam optimizer with a learning rate of $2 \times 10^{-5}$ and cosine annealing, for 20 epochs with early stopping.

We defined four spoofing class labels:

\begin{itemize}
    \item \textbf{0 – Human:} Authentic speech recorded from real speakers.
    \item \textbf{1 – Cloned:} Speech synthesized with Tacotron2 using speaker embeddings.
    \item \textbf{2 – AI-generated:} Fully synthetic speech without speaker identity preservation.
    \item \textbf{3 – Hybrid:} Concatenated sequences mixing human and synthetic segments.
\end{itemize}

This fine-tuning pipeline bridges pretrained vision knowledge with task-specific audio spoof detection, achieving both high classification accuracy and generalization to noisy and adversarial inputs.

%% file: 05-Experiment.tex
\vspace{-0.2cm}
\section{Evaluation}

\subsection{Experimental Setup}

The evaluation was conducted using both clean and noisy versions of the proposed hybrid spoofing dataset. Four spoofing classes were defined: (0) genuine human speech, (1) AI-cloned speech, (2) AI-generated speech, and (3) hybrid (mixed-source) speech. These classes reflect realistic adversarial scenarios designed to challenge anti-spoofing models under varied linguistic, acoustic, and synthesis conditions.

Audio recordings were sampled at 16 kHz and converted into 128-bin log-Mel spectrograms using a 25 ms Hamming window and a 10 ms frame shift. All experiments were conducted on an NVIDIA A100 GPU. Training followed an 80/20 train-test split, ensuring speaker disjointness to prevent overfitting.

Each model was trained for 20 epochs using the Adam optimizer with an initial learning rate of $2 \times 10^{-5}$, cosine decay scheduling, and early stopping. Batch size was fixed at 3 for stability given the variable input lengths and memory constraints.

\subsection{Evaluation Metrics}

Performance was measured using classification accuracy, F1-score, false positive rate (FPR), and false negative rate (FNR) across the four spoofing classes. For reliability-based binary classification (used in baseline benchmarking), the following thresholding rule was applied:

\[
\hat{y} =
\begin{cases}
0 & \text{if } |\text{real\_tag} - \text{reliability\_score}| < 0.5 \\
1 & \text{otherwise}
\end{cases}
\]

Overall classification accuracy was computed as:

\[
\text{Accuracy} = \frac{C}{N} \times 100\%
\]

where \(C\) is the number of correct predictions and \(N\) is the total number of samples. Confusion matrices and reliability score distributions were used to analyze error patterns and inter-class ambiguities.

\subsection{Datasets}

To establish robust benchmark performance for spoofed speech detection, we conducted evaluations on two datasets: the widely used ASVspoof 2019 Logical Access (LA) dataset~\cite{Vestman2019} and our newly constructed hybrid spoofed audio dataset, introduced in Section~\ref{sec:dataset-construction}.

\textbf{ASVspoof 2019 LA Dataset.} The ASVspoof 2019 LA dataset serves as a standard benchmark in the field of automatic speaker verification (ASV). It comprises bona fide and spoofed utterances generated using advanced text-to-speech (TTS) and voice conversion (VC) algorithms. The dataset is organized into disjoint training, development, and evaluation sets, offering a comprehensive framework to assess generalization under realistic synthesis conditions.

\textbf{Proposed Hybrid Spoofed Audio Dataset (HSAD).}  
To overcome the limitations of existing benchmarks—such as binary classification focus, limited spoof type diversity, and the absence of hybrid composition realism—we constructed the HSAD dataset. It includes six carefully designed categories that reflect a spectrum of real-world and adversarial audio conditions:

\begin{itemize}
    \item \textbf{G1 – Genuine Human:} Natural, untouched speech recordings from real speakers.
    \item \textbf{G2 – Pure AI Clone:} Cloned speech generated using Tacotron 2-based models with varying levels of speaker reference embeddings (e.g., single-sentence, corpus-level, and target-matched).
    \item \textbf{G3 – Pure AI Generated:} Fully synthetic speech created via zero-shot TTS systems with no prior speaker conditioning.
    \item \textbf{G4 – Mixed: AI Generated + Human:} Spliced utterances formed by concatenating segments of AI-generated and human speech to simulate content injection attacks.
    \item \textbf{G5 – Mixed: AI Cloned + AI Generated:} Utterances combining cloned and AI-generated segments, introducing complex spoofing configurations.
    \item \textbf{G6 – Human Recombined:} Human-only segments rearranged into hybrid-like flows to emulate natural conversational variation without introducing spoof artifacts.
\end{itemize}

Each audio sample is labeled with speaker ID, spoof type, segment structure, and signal fidelity level. Beyond clean speech, the dataset also includes an adversarial variant with environmental noise (10–30 dB SNR), channel filtering (low-pass at 4 kHz), and compression artifacts (Opus codec at 16–24 kbps) to simulate deployment conditions. These features support detailed robustness evaluation and promote generalization studies for both fine-grained classification and real-world ASV defense mechanisms.

\subsection{Baseline Models}

To evaluate performance on these datasets, we selected three state-of-the-art transformer-based models from the Hugging Face model repository:

\textbf{MIT-AST}~\cite{MITast}: A general-purpose Audio Spectrogram Transformer (AST) model pretrained on the AudioSet corpus using weakly labeled 10-second audio segments. This model captures diverse acoustic events and serves as a strong baseline for general audio classification tasks.

\textbf{MattyB95}~\cite{MattyB95ASTASVspoof5}: A domain-specific AST model fine-tuned directly on the ASVspoof 2019 LA dataset for detecting synthetic speech. It provides a strong reference for binary spoof detection.

\textbf{WpythonW}~\cite{WpythonW}: This model is trained on ElevenLabs synthetic speech and provides an additional comparison point, particularly for evaluating generalization across spoof generation techniques not included in the ASVspoof dataset.

Together, these models offer diverse architectural and training configurations—ranging from broad-spectrum audio classification to specialized spoofing detection—enabling comprehensive evaluation of detection capabilities under both standard and newly proposed conditions.

\subsection{Performance on ASVspoof 2019 LA}

To evaluate baseline spoof detection performance, we tested three transformer-based models on the ASVspoof 2019 Logical Access (LA) dataset~\cite{Vestman2019}. These included MIT-AST, a general-purpose Audio Spectrogram Transformer (AST) model pretrained on AudioSet; MattyB95, which was specifically fine-tuned on ASVspoof 2019 challenge data; and WpythonW, a variant trained with ElevenLabs-generated synthetic speech. Table~\ref{tab:model_performance} presents the number of correct predictions and overall accuracy for each model.

\begin{table}[t]
    \caption{Performance on ASVspoof 2019 LA Dataset.}
			\vspace{-0.2cm}
	\centering
	\small
    
	\setlength{\extrarowheight}{0.5pt}
	\begin{tabular}
		{m{1.7cm}<{\centering} m{2.2cm}<{\centering} m{1.9cm}<{\centering}} 
		\Xhline{1.5\arrayrulewidth}
		\textbf{Model Name} &\textbf{Correct Predictions} &\begin{tabular}{@{}c@{}}\textbf{Accuracy ( \%)}\end{tabular}\\
		\Xhline{1.5\arrayrulewidth}
        MIT-AST & 63,663 / 71,237 & 89.37\% \\
        MattyB95 & 63,863 / 71,237 & 89.65\% \\
        WpythonW & 42,143 / 71,237 & 59.16\% \\
		\Xhline{1.5\arrayrulewidth}
	\end{tabular}
		\vspace{-0.4cm}
	\label{table:datasets}
\end{table}
\vspace{-0.1cm}

The results confirm that transformer-based architectures are highly effective for speech spoof detection. Both the MIT and MattyB95 models achieved nearly identical performance, with accuracy rates exceeding 89\%, demonstrating that large-scale audio pretraining (e.g., on AudioSet) and fine-tuning on task-specific datasets both contribute significantly to model robustness. Notably, MattyB95, which was explicitly optimized on ASVspoof 2019, achieved the highest accuracy, albeit with only a marginal improvement over MIT’s general-purpose variant. This highlights the value of domain-aligned tuning for maximizing performance on known spoofing types.

In contrast, the WpythonW model exhibited significantly lower accuracy at just 59.16\%, despite being based on the same architectural backbone. This substantial performance drop indicates that models trained narrowly on specific synthetic speech generators may suffer from poor generalization when exposed to a wider variety of spoofed content. The reliance on ElevenLabs speech alone likely contributed to overfitting, making the model less effective on broader spoof distributions present in the ASVspoof benchmark.

These findings collectively underscore the importance of balancing large-scale, diverse pretraining with targeted fine-tuning. Pretrained transformer models provide a strong foundation, but their ability to detect spoofed content across diverse conditions depends critically on exposure to representative spoofing variations during the fine-tuning phase. Future evaluation metrics should move beyond raw accuracy to consider generalization across spoofing techniques and robustness under real-world signal degradation, which will be explored in subsequent sections.

\subsection{Performance on the Proposed HSAD Dataset}

To evaluate the robustness and generalization capabilities of transformer-based models under realistic spoofing conditions, we assessed the two best-performing baseline models—MIT-AST and MattyB95—on our proposed Hybrid Spoofed Audio Detection (HSAD) dataset.

Table~\ref{table:datasets-1} summarizes the reliability score statistics for both models across six spoofing groups: \textbf{G1} (Genuine Human), \textbf{G2} (Pure AI Clone), \textbf{G3} (Pure AI Generated), \textbf{G4} (Mixed: AI Generated + Human), \textbf{G5} (Mixed: AI Clone + AI Generated), and \textbf{G6} (Human Recombined).

\begin{table}[t]
\caption{Reliability score statistics for six spoofing groups (G1–G6) evaluated using MIT-AST and MattyB95 models on the proposed HSAD dataset.}
\vspace{-0.3cm}
\centering
\small
\begin{tabular}{m{0.2cm}<{\centering} m{0.7cm}<{\centering} m{1.0cm}<{\centering} m{0.7cm}<{\centering} m{0.7cm}<{\centering} m{0.9cm}<{\centering} m{0.9cm}<{\centering}}
\Xhline{1.5\arrayrulewidth}
\quad & \textbf{Group} & \textbf{Mean} & \textbf{Std Dev} & \textbf{Max} & \textbf{Min} & \textbf{Mode} \\
\Xhline{1.5\arrayrulewidth}
\multirow{6}*{\rotatebox{90}{MIT-AST}} 
& G1 & 0.8414 & 0.1339 & 0.9768 & 0.0071 & 0.7977 \\ 
& G2 & 0.8756 & 0.0829 & 0.9923 & 0.0185 & 0.8828 \\ 
& G3 & 0.8606 & 0.0959 & 0.9639 & 0.3924 & 0.7320 \\ 
& G4 & 0.8009 & 0.1398 & 0.9827 & 0.0908 & 0.7479 \\ 
& G5 & 0.8092 & 0.1343 & 0.9927 & 0.0185 & 0.7705 \\
& G6 & 0.7904 & 0.1748 & 0.9794 & 0.0095 & 0.0095 \\ 
\Xhline{1.0\arrayrulewidth}
\multirow{6}*{\rotatebox{90}{MattyB95}} 
& G1 & 0.5385 & 0.4925 & 1.0000 & 5.96e-7 & 5.96e-7 \\
& G2 & 0.9821 & 0.1319 & 1.0000 & 5.96e-7 & 1.0000 \\ 
& G3 & 0.3519 & 0.3519 & 1.0000 & 5.96e-7 & 5.96e-7 \\ 
& G4 & 0.3669 & 0.4733 & 1.0000 & 5.96e-7 & 5.96e-7 \\ 
& G5 & 0.6658 & 0.4679 & 1.0000 & 5.96e-7 & 5.96e-7 \\ 
& G6 & 0.2713 & 0.4435 & 1.0000 & 5.96e-7 & 5.96e-7 \\
\Xhline{1.5\arrayrulewidth}
\end{tabular}
\label{table:datasets-1}
\vspace{-0.4cm}
\end{table}

The MIT-AST model, despite achieving a high average accuracy of 93.67\% on the HSAD dataset, failed to correctly identify any genuine human utterances, misclassifying them as spoofed. This is evidenced by the close overlap of mean reliability scores for Group G1 (Human, 0.8414), G2 (Cloned, 0.8756), and G3 (Generated, 0.8606). Such overlap indicates insufficient class separation and undermines model interpretability and trustworthiness in real-world deployment.

In contrast, the MattyB95 model exhibited a lower overall accuracy of 65\% but showed improved class distinction. Its reliability scores for genuine human speech (G1: 0.5385) and human recombined segments (G6: 0.2713) were substantially lower than those for cloned (G2: 0.9821) and hybrid compositions, offering better separation between genuine and synthetic content.

However, both models struggled with the hybrid categories. Groups G4 (AI Generated + Human) and G5 (AI Clone + AI Generated) yielded highly dispersed scores with large standard deviations, suggesting confusion due to complex boundary conditions and mixed-source signal characteristics.

These results underscore three critical insights:

1. **Limitations of Standard Pretraining:** Models like MIT-AST, despite broad pretraining on AudioSet, are not calibrated for fine-grained spoof discrimination and tend to overgeneralize, especially in the presence of hybrid or partially spoofed audio.

2. **Benefit of Spoof-Specific Tuning:** While MattyB95 exhibits better discrimination between real and spoofed speech, its performance degrades under distribution shifts, such as unseen hybrid constructs.

3. **HSAD Dataset Utility:** The HSAD dataset introduces nuanced scenarios and spoofing combinations absent from traditional corpora, highlighting its importance in benchmarking robust, future-ready detection architectures.

In conclusion, existing models fail to reliably detect and classify hybrid spoofing attacks due to overlapping decision boundaries. These findings motivate the development of hybrid-aware models with finer temporal segmentation, semantic consistency modeling, and adaptive spoof calibration strategies tailored to composite real-world conditions.

\subsection{Fine-Tuned Model Performance}

To validate the efficacy of the proposed HSAD dataset, we fine-tuned two transformer-based models—Model A, adapted from the MIT-AST architecture, and Model B, derived from MattyB95. Both models retained the Audio Spectrogram Transformer (AST) backbone and were retrained on our multi-source dataset to enhance domain-specific spoof detection.

Training was conducted for 20 epochs using the Adam optimizer and cosine learning rate scheduling, with early stopping based on validation loss. Audio inputs were standardized to 128-bin log-Mel spectrograms computed with 25ms frame length and 10ms hop size. A batch size of 3 was used, and training was performed on an NVIDIA A100 GPU.

Model A achieved a validation accuracy of 97.88\%, while Model B slightly exceeded this with 98.08\%. On the test set, both models stabilized at 97\% accuracy, substantially outperforming the baseline AST variants. The confusion matrices illustrate these improvements. Model A (fine-tuned MIT-AST) achieved particularly consistent performance across all categories, while Model B (fine-tuned MattyB95) showed strong discriminability among hybrid and synthetic speech types.

\begin{figure}[H]
    \centering
    \includegraphics[width=0.5\linewidth]{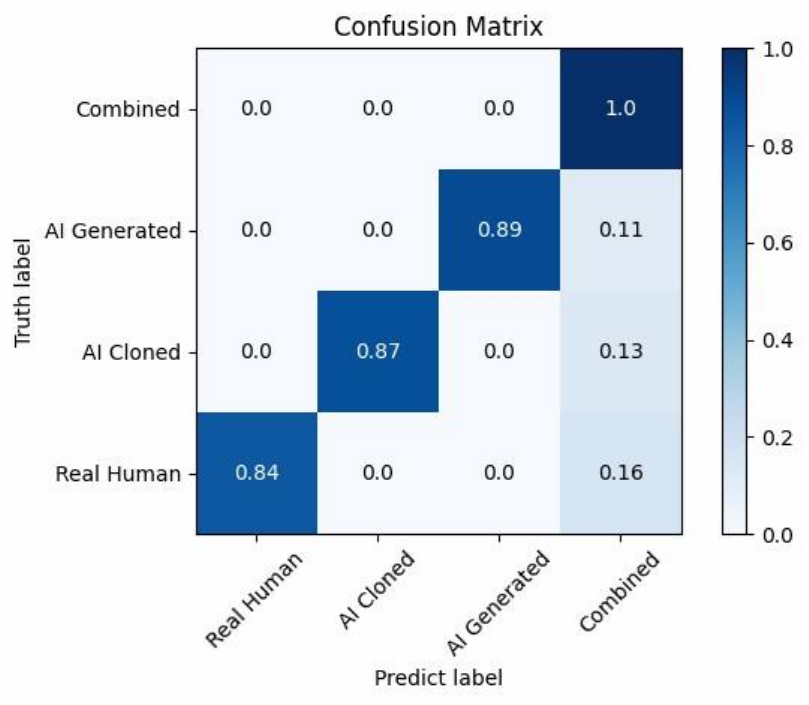}
    \caption{Confusion matrix – MIT fine-tuned (Model A)}
\end{figure}
\begin{figure}[H]
    \centering
    \includegraphics[width=0.5\linewidth]{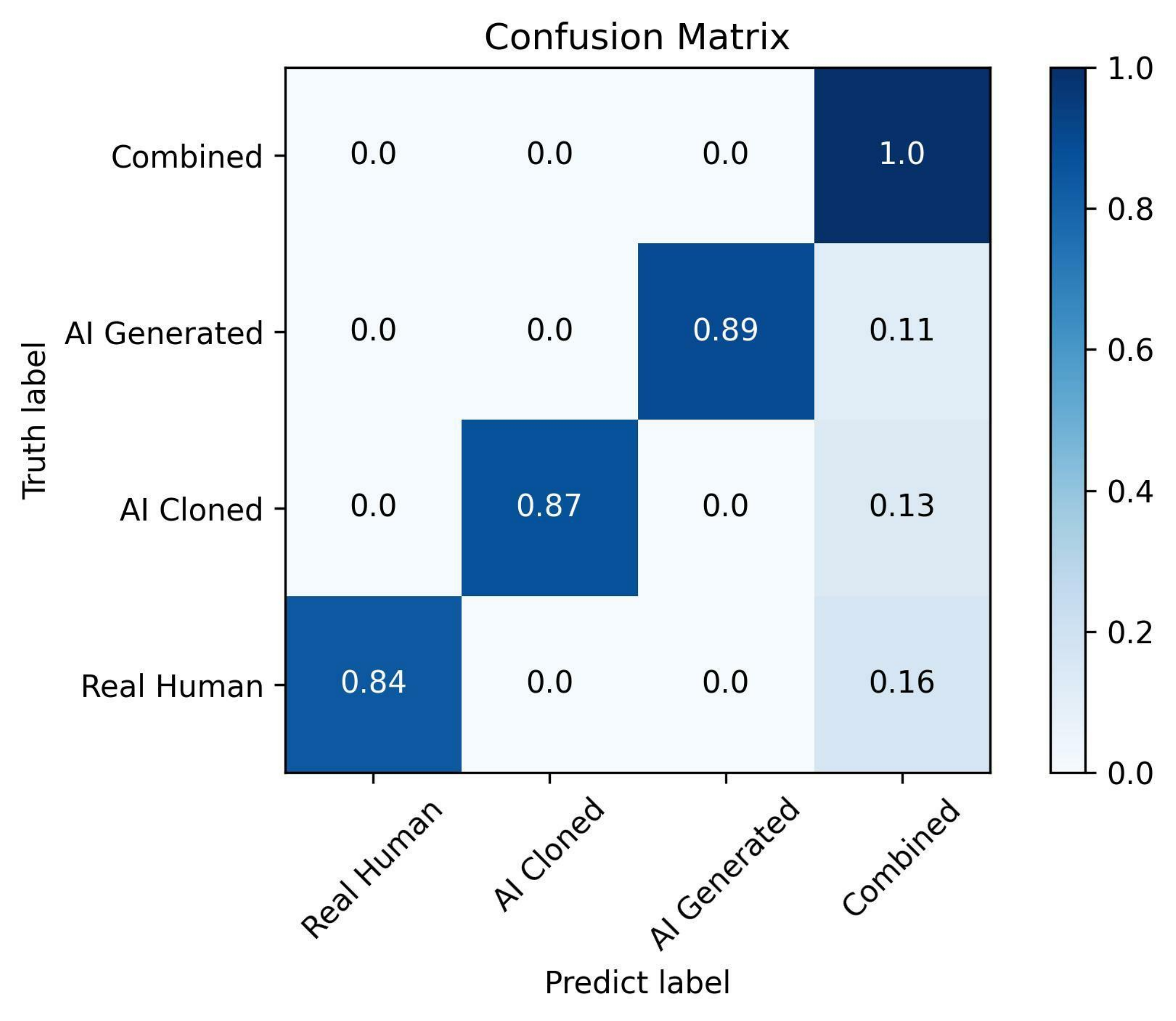}
    \caption{Confusion matrix – MattyB95 fine-tuned (Model B)}
\end{figure}

Both models correctly identified hybrid audio samples with 100\% precision, demonstrating their ability to capture compositional cues that standard classifiers often miss. For AI-generated and AI-cloned content, recall exceeded 88\%, confirming their capability to recognize nuanced synthetic speech even when integrated into natural audio flows. The classification accuracy for human speech rose to 84\%, representing a dramatic improvement compared to the baseline MIT-AST model, which had previously misclassified all genuine utterances.

\begin{figure}[H]
    \centering
    \includegraphics[width=\linewidth]{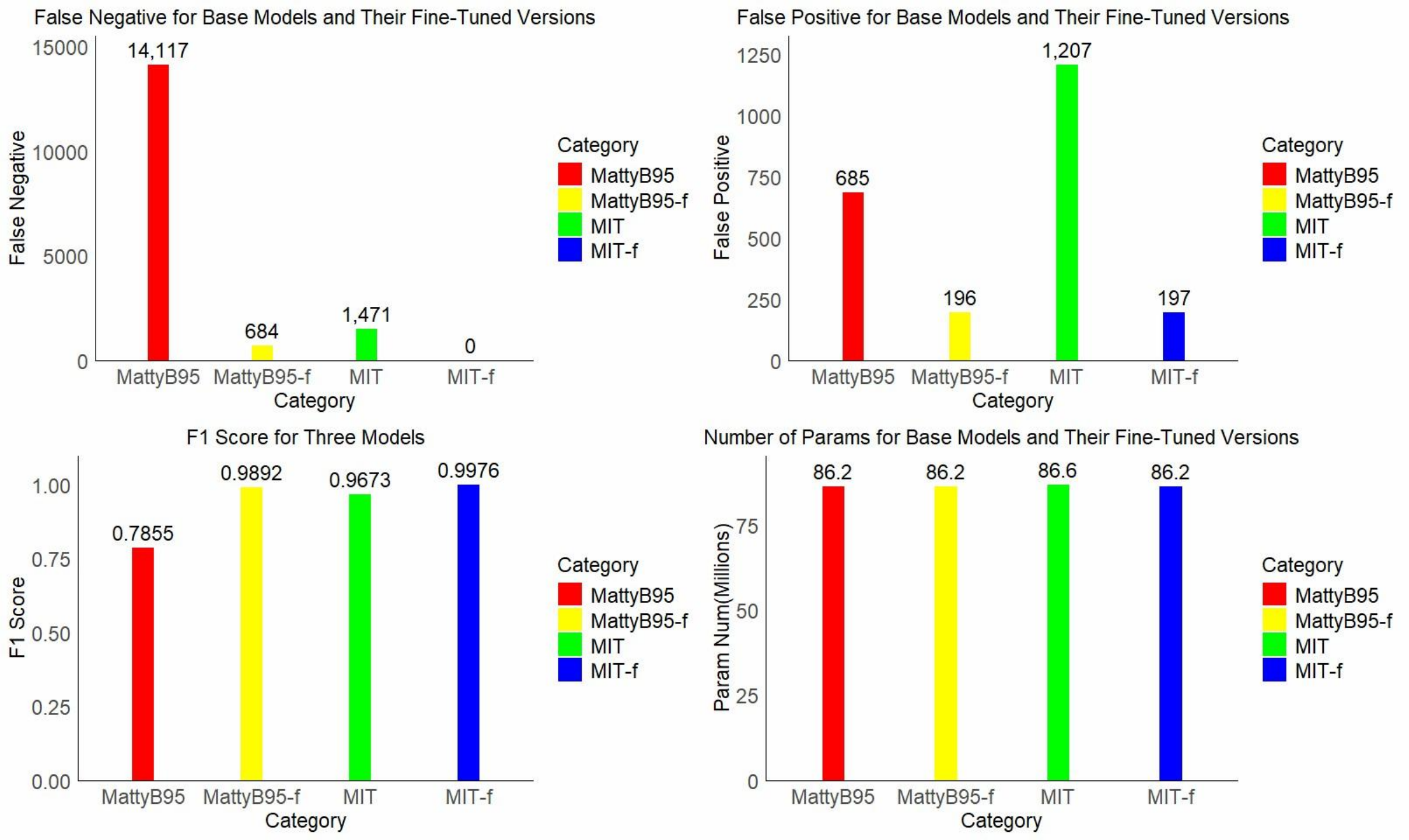}
    \caption{Comparison of false positives, false negatives, F1-score, and parameter count across baseline and fine-tuned models}
    \label{fig:all_comb}
\end{figure}

Performance gains extended beyond accuracy. As shown in Fig.~\ref{fig:all_comb}, Model A achieved a 71\% reduction in false positives for human speech (from 1,207 to 197), while Model B demonstrated a 95\% reduction in false negatives for AI-generated content (from 14,117 to 684). Both models reached an F1-score of 99\%, indicating strong balance between precision and recall. Furthermore, Model A reduced its parameter count by 0.4M compared to the original MIT-AST (from 86.6M to 86.2M) while achieving a 2.27\% F1-score gain, highlighting the efficiency and effectiveness of dataset-specific adaptation.

These findings emphasize that models pretrained on general-purpose audio datasets struggle with complex spoof compositions unless retrained on spoof-specific corpora. Although MIT-AST benefited from broad AudioSet pretraining, it lacked calibration for spoof detection tasks, particularly under hybrid scenarios. Conversely, the MattyB95 baseline model exhibited a stronger spoof detection bias but suffered from low reliability scores and misclassifications on unfamiliar or subtle combinations of synthetic and real audio. After fine-tuning, both models demonstrated significant improvements in generalization, particularly under adversarial and mixed-source conditions.

In conclusion, fine-tuning on the HSAD dataset not only improved classification accuracy and F1-score, but also enhanced the models’ robustness against false classifications and improved computational efficiency. These results affirm the importance of dataset-specific training for spoof detection and underline the potential of HSAD as a benchmark for advancing robust, real-world anti-spoofing solutions.

\subsection{Discussion}

This study introduces a systematically constructed Hybrid and Spoofed Audio Dataset (HSAD) designed to expose the limitations of current state-of-the-art anti-spoofing systems under complex and adversarial conditions. Unlike existing corpora such as ASVspoof 2019, which focus predominantly on binary classification and clean synthetic speech, our dataset reflects real-world threat vectors through four spoofing classes: genuine human speech, AI-cloned speech, fully AI-generated speech, and various hybrid compositions combining human and synthetic audio segments. Additionally, we inject practical distortions such as environmental noise, codec compression, and channel degradation to simulate mobile and cross-platform transmission conditions.

Empirical results from this work highlight a significant performance gap between models trained on homogeneous public datasets and their effectiveness when deployed in adversarial environments. Although the MIT/AudioSet transformer model achieved over 93\% accuracy in aggregate, it misclassified all genuine human samples as spoofed, revealing a critical flaw in model calibration. The MattyB95 baseline, despite being fine-tuned on the ASVspoof 2019 dataset, showed improved class separability but struggled with mixed-source inputs due to overlapping reliability distributions.

In contrast, fine-tuned models trained directly on HSAD achieved substantially higher performance across all metrics. With accuracy exceeding 97\% and F1-scores approaching 99\%, these models reduced false positive rates for human speech by over 70\% and false negatives for synthetic speech by over 90\%. These results not only validate the advantage of dataset-specific fine-tuning, but also emphasize the importance of including hybridized spoof structures during training to account for increasingly subtle and compositional spoofing techniques.






\subsection{Limitations and Future Works}
\subsubsection{Limitations}

Despite the substantial contributions of this work, several limitations must be acknowledged to contextualize the findings:

\textbf{1. Spoofing Coverage.} The proposed HSAD dataset currently focuses on two primary spoofing paradigms: text-to-speech (TTS) synthesis and AI-based voice cloning. It does not yet incorporate other advanced attack vectors such as voice conversion (VC), prosody manipulation, GAN-generated speech, or adversarial examples designed to evade detection systems. This restricts the coverage of spoofing techniques and may limit model robustness under more exotic or unseen attacks.

\textbf{2. Environmental Realism.} Although HSAD includes artificial distortions such as additive noise, codec compression, and filtering, all clean speech recordings were collected under controlled acoustic environments. Realistic variability—including spontaneous speech, reverberant rooms, overlapping speakers, mobile device microphones, and network-based degradation—remains unaddressed. This may introduce evaluation bias when deploying models in truly unconstrained conditions.

\textbf{3. Dataset Scale and Generalization.} The speaker pool is limited to twelve individuals, which, although balanced in terms of age and gender, may not reflect broader phonetic diversity. Moreover, the number of utterances per spoofing category is constrained. Consequently, there is a risk of overfitting to the dataset-specific patterns, especially when models are deployed on cross-domain benchmarks or real-world applications.

\textbf{4. Hybrid Annotation Granularity.} While the dataset includes hybrid compositions, annotations are currently coarse-grained—focusing on utterance-level authenticity rather than fine-grained localization of spoofed segments. This limits the ability to analyze and interpret partial spoofing strategies or perform segment-level detection.

\subsubsection{Future Work}

Building on the current work, several directions are planned to further improve dataset realism, model robustness, and interpretability:

\textbf{1. Expansion of Spoof Modalities.} Future iterations of HSAD will include broader spoofing types such as voice conversion, neural vocoder-based attacks, multilingual TTS, and adversarial attacks crafted using gradient-based or black-box methods. This will improve the comprehensiveness of the dataset and increase the challenge for detection models.

\textbf{2. Real-World Data Collection.} To better approximate deployment scenarios, we will incorporate audio collected in the wild using smartphones, smart speakers, and telephony-grade systems. Data will span multiple environments (e.g., urban outdoors, homes, transit), capturing the variability encountered in everyday usage.

\textbf{3. Segment-Level Annotation and Localization.} Future releases of HSAD will include precise time-stamped annotations for the onset and offset of synthetic segments within hybrid utterances. This will enable the training and evaluation of spoof localization models, and support applications such as real-time detection and signal repair.

\textbf{4. Cross-Corpus Benchmarking.} We will benchmark our models on additional datasets such as ADD2023-PF, Half-and-Half (HAD), and multilingual corpora to evaluate cross-domain transferability. This will help establish HSAD’s utility as a training and diagnostic resource for general-purpose anti-spoofing systems.

\textbf{5. Model Explainability and Robustness Analysis.} We plan to integrate explainable AI (XAI) techniques such as attention heatmaps and gradient-based saliency to better understand model decisions, especially under ambiguous or adversarial inputs. Robustness evaluation under perturbation (e.g., adversarial noise, pitch shifting) will also be explored.

\noindent In conclusion, while HSAD and its associated fine-tuned models represent a significant advance in spoof detection, the field remains dynamic. Continued expansion of spoofing strategies, environmental realism, and explainability mechanisms will be critical for developing trustworthy, real-world voice security systems.

%% file: 06-Conclusion.tex
\section{Conclusion}

This study presented the Hybrid Spoofed Audio Dataset (HSAD), a novel and realistic benchmark aimed at advancing the robustness of automatic speaker verification (ASV) systems against diverse and hybrid spoofing threats. Unlike prior corpora limited to binary spoof labels, HSAD introduces six fine-grained spoof categories—including AI-cloned, zero-shot generated, and mixed-source utterances—constructed via Tacotron-based voice cloning, zero-shot synthesis, and hybrid segment splicing. Real-world conditions were further simulated through noise injection, codec compression, and spectral filtering.

Comprehensive evaluations using two state-of-the-art Audio Spectrogram Transformer (AST) models pretrained on AudioSet and ASVspoof 2019 revealed critical limitations. While the MIT-AST model achieved 93.67\% overall accuracy, it failed to correctly identify genuine human audio. The MattyB95 model performed better on human detection but suffered from reduced confidence and poor separation across hybrid categories. These results highlight the poor generalization of models trained on traditional, homogeneous spoof datasets.

Fine-tuning both models on HSAD led to substantial performance gains—elevating classification accuracy to 97\%, boosting F1-scores to 99\%, and dramatically reducing false positive and false negative rates. This underscores the necessity of dataset-specific adaptation and the value of richly annotated, diverse spoof datasets for training spoof-aware detection models.

Looking ahead, future work will expand the scope of spoof types to include voice conversion (VC), adversarially crafted attacks, and multilingual synthesis. Real-world recordings captured through mobile and conferencing devices will further increase environmental realism. Finally, explainable AI (XAI) methods will be integrated to support interpretability and temporal spoof localization within composite audio streams.

In summary, HSAD establishes a new foundation for anti-spoofing research, addressing key gaps in existing benchmarks and demonstrating the transformative effect of task-aligned fine-tuning for transformer-based ASV systems. This work paves the way toward more secure and trustworthy audio verification systems in real-world applications.